# On Pattern and Evolution
F. W. Cummings*

**Abstract**
A theoretical model of biological patterning consisting of two determined regions interspersed by a smaller third region is presented. This patterning is not dependent on long range diffusion, but only on short range or near neighbor diffusion. The interaction of two signaling pathways is discussed as a general basis of such tripartite biological patterning. Plant patterns (phyllotaxis) illustrate the model without the complicating factor of shape changes; point-like regions of outgrowth from a stalk are specified on a cylindrical geometry. This signal pathway interaction is further proposed as a key component in the transition from single-celled to multicellular life. Gastrulation is obtained numerically as a coupling of the pattern model to geometrical change; pattern changes geometry via the genes, and geometry in turn changes pattern. Contact of the animal and vegetal poles starting from a gastrulating sphere is the starting point for examining simple conditions giving bifurcation into 'Urcnidaria' and 'Urbilateria'. Pattern specification of "master regulatory genes" (e.g., pax-6, distal-less) at specific 'points' is suggested. Such positional specification emerges at intersection(s) of two 'thin lines' of the smaller interstitial regions dividing the two main determined (e.g., by Hh or Wnt) regions, as in the plant patterns.

Key words: evolution, signaling pathways, pattern, plant patterns

*Professor emeritus, University of California Riverside.
Present address, to which all correspondence should be addressed:
136 Calumet Ave., San Anselmo, Ca., 94960.
fredcmgs@berkeley.edu .

# 1. Introduction

It has now become clear that virtually all developmental regulatory genes control several different processes, acquiring new developmental roles. Clusters of Hox genes, as well as Pax-6, Dll and Tinman proteins, along with many others, shape the development of animals as diverse as flies and mice. These genes and their proteins are just a part of the collection that make up the genetic 'tool kit' for animal development. Transcription factors are proteins that bind to DNA and directly turn gene transcription on or off, and comprise a large fraction of the regulatory tool kit. The present view is that although developmental regulatory genes are remarkably conserved, their interactions are not (Carroll et al., 2001; Davison, 2001; Wilkins, 2002; Carroll, 2005). The recent impressive progress in unraveling the genomic basis of development and evolution has led to a great advance in understanding of animal development.

However, what is further clear is that elucidation of the actual cell shape changes along with an understanding of the causes of changes in tissue shape and cell number is necessary to obtain a fuller grasp of morphogenesis. Further, such issues as the positioning (patterning) of stem cells, as well as designation of precise positioning of eyes, antennae, legs, wings, gills remain to be addressed by a patterning model.

Communication between cells must play a decisive role in development. Natural selection is the principal influence driving evolution. The argument here is that, acting along with natural selection, are generative 'rules', from the very origin of multicellularity, which lead to bias or constraint on natural selection. A number of authors have previously argued that this is the case (Arthur, 2002; Webster and Goodwin, 1996; J.M. Smith et al., 1985). It seems possible or even probable that remnants of the action of such constraint remain today, even after the extensive elaboration of more than 600 million years of evolution. Such elaboration upon the primitive rules would suggest that while such rules have become obscured, they may be still accessible.

Thus we argue that sophisticated eukaryotic cells found a way to form multicellulars before the Cambrian, discovering 'rules' that were 'adaptive' at that time, and although extensively elaborated since then, have left evidence as to their form and origin. Of interest here is to propose a possible key component in the origin of the metazoan, one still operative today. The rules or model proposed in what follows are no doubt too simplified to be realistic, but it is hoped that they will introduce less familiar concepts, and will act to stimulate further investigation along these lines.

It has become customary to assume that patterning occurs by way of small diffusible molecules. These are assumed to diffuse over long enough distances (say, thirty cells) so that cells at different positions in the resultant gradient somehow read the concentration of the diffusible molecules, allowing cells to determine patterns of gene expression. Such diffusible 'morphogens' have long been the standard framework for interpretation of experiments involving pattern formation (Kerszberg and Wolpert, 2007). In fact, the term 'morphogen' has even come to specify such long range diffusible molecules, basing the meaning of the very term on a particular model, rather than on its Greek root meaning "shape genesis".

The gradient model has recently come under careful and sharp experimental criticism; at best, such gradients are not the whole story (Gregor et al., 2007; Kerszberg and Wolpert, 1998, 2007; Kornberg and Guha, 2007). It is argued in the present work that long range diffusion is even unnecessary for pattern formation, and only short range diffusion, as short perhaps as nearest neighbor diffusion, can establish most needed patterns. Many of the problems related to morphogen propagation and gradient establishment were recognized long ago by Wilson and Melton (1994).

Among the necessary information that patterns must supply, cells must be able to acquire gradient specification in the plane of the epithelium, that is, acquire planar polarity. Hair cells of the fly wing, for example, are polarized in the plane of the epithelium (Lawrence et al., 2004). Another demanding condition is that boundaries separating determined regions may be specified within a single cell diameter, occurring perhaps during insect segmentation at the antero–posterior parasegment boundaries. Cell-cell interactions are then strongly indicated as essential players (Kerzberg and Wolpert, 2007). Also, growth must be included as an integral Aspect of a reasonable patterning model.

Further, Kornberg and Guha (Kornberg and Guha, 2007) have argued convincingly (on the basis of experiments on fly wing imaginal discs) that in the absence of constraining impenetrable physical barriers, gradient-generating dispersion of morphogens cannot be achieved by passive long range diffusion. This method of patterning is not compatible with the need to pattern different regions separated to within a single cell diameter.

The present work is divided into several sections. Section 2 develops the basic tripartite pattern model as arising from the interaction of two signaling pathways. The patterns may arise spontaneously, from homogeneous origins, 'Turing like', at each growth cycle. With growth, two different signaling pathways may subsequently become active following the previous cycle, and the previous pathways may become deactivated after having achieving the desired genetic determination.

A relatively thin region will separate two regions of genetic determination in each growth cycle. We will here designate this thin region as a '**Margin**' region, and the cells in it **'Margin cells'** ( Figure 1). The term **'Asp'** on the other hand will denote an "activated signalling pathway", a region of activation of a particular transcription factor, such as that of Wnt or Hedgehog (Hh). Two different **Asp** regions will border or enclose the smaller, usually line-like **Margin** region. A tripartite pattern thus emerges. With this designation, we hope to avoid collision with the present usage of the word "morphogen" by biologists as a 'diffusing substance'.

Section 3 deals with the development of the earliest metazoans. It is argued that adhesive cell-cell connections, and the cell-cell communication and patterning afforded by two coupled signaling pathways (esp., 'Wnt' and Hedgehog ('Hh') (Duman-Scheel, 2002; Willert et al., 2003; Garcia-Castro et al., 2002; Zhao, Tong and Jiang, 2007) leads naturally to gastrulation. Gastrulation, the formation of a "tube within a tube", is a most adaptive and universal feature of animals. The epithelial deformations involved in gastrulation, starting from the spherical blastula, are of fundamental interest.

Section 3 goes beyond the axially symmetric patterns of 'proto-gastrula'. Again, only the earliest (~ Cambrian) and most primitive organisms are of interest. Only simple axially symmetric geometrical forms are examined. It is expected that patterns on such simple geometries will provide basic insight into more geometrically realistic situations.

In particular, solutions are found on a (thin) cylinder within a cylinder, mimicking a donut (a hollow donut) extended along its axis, a torus Analytic solutions are not easily obtained in non-axially symmetric solutions of patterns coupled to geometrical changes, and such computationally difficult solutions are not attempted here.

Formation of a through-gut animal follows gastrulation, protostomes or deuterostomes, when the animal and vegetal poles touch, and cell death takes place. The situation where segments are added at each cycle as the animal grows is considered, mimicking the way certain "short germ" arthropod multicellulars were able to form, starting from a single cells. Growth occurs from the most posterior **Margin region**. Growth along the body axis occurs in the posterior region of very many animals, and the Hox genes are in the same order along the chromosome as along the anterior-posterior animal axis, in agreement with this picture.

The spatial patterning of projections from the main body of many bilateral animals, initiated by regulatory proteins such as Distal-less ('Dll'), and including appendages such as limbs, gills, or wings, are specified in the model as a point-like region by two intersecting '**Margin cell**', line-like regions.

Outgrowth from the main body axis of both animals and plants is accurately patterned by the same basic mechanism in the present model; outgrowth occurs at points specified by two intersecting 'Margin cell' lines. Lines are designated as the middle of a margin cell region. Such 'points' of intersection are also speculated to activate "master regulatory" genes at antennae, eyes, wings, and gill positions. It is of some interest to notice that only an odd number of leg pairs appear if it is presumed that they emanate from the intersections of Margin regions provided by the model, when the assumption is that a single Hox cluster does not differentiate segments. This 'odd-pair' phenomenon occurs in the over 3,000 species of centipede examined. Such does not allow explanation by natural selection.

Section 4 deals with 'leaf' patterning on a (cylindrical) plant stem. The patterning here further illustrates the patterning capability of the proposed model. All outgrowths ('leafs') from a stalk occur at intersection of two Margin regions, as in animals. All leaf patterns are reproduced by a simple algorithm. Two new and smaller integers (p, q) replace the more familiar ones (m, n). It is shown that the number 'four' does not occur in a spiral pattern with one 'leaf' per level, according to both the model and observation. All plant outgrowths are assumed to be patterned by the same general model as in the animal case, although of course, the biochemicals involved are expected to be quite different; 'aux' appears to play a crucial role in plants, similar to that played by Wnt or Hh in the animal case. The points of outgrowth from the stalk occurs as in animals, namely at places specified by the intersection of two 'Margin cell' line

A prediction is that a universal plant analogue to the 'master regulatory' distal-less (dll) gene of animal trunk outgrowth will be found.

**2. The interaction of two signaling pathways and the origin of multicellularity**

The origin of multicellular life from single-celled beginnings is one of the most enigmatic of puzzles, and one least likely to be ever 'solved'. However, the question will continue to exercise the imagination, as it has for hundreds of years. The multiplicity of theories concerning the origin of multicellularity has been ably reviewed (Willmer, 1990). The work of Haeckel (~1874) on metazoan origins involving blastula-like and gastrula-like stages has long been influential, and since that time there have been numerous alternate proposals (Willmer, 1990; Valentine, 2004). Haeckel's ideas, while having many virtues such as simplicity, elegance and orderliness, are open to several objections.

Often a zooflagellate colony similar to Volvox is invoked as the earliest ancestor of multicellulars. In this view, the basic metazoan was a pelagic, radially symmetric aggregation of flagellated cells. Such an aggregation of cells has several desired properties, such as a separation between somatic and gametic cells, a blastula-like geometry, and cells having loose connections at their lateral surfaces. There are many species of colonial flagellate protists. Recent findings (Abedin and King, 2008) as well as the discussion of Nielsen (Nielsen, 2001) offer a convincing origin of multicellulars as being from the single-celled choanoflagellates.

A key assumption of the present work is that a crucial 'discovery' by evolution at or near the critical turning point leading to multicellulars was of the patterning potential of the interaction of two signaling pathways. Needless to say, concurrent felicitous environmental conditions, as well as complex cellular development would be necessary preconditions, but are not discussed further here.

The simplest patterning mechanism is proposed in this section. Several elements are essential in any discussion of the origin of multicellulars from single celled eukaryotes. The first is the development of adhesive connections between cells; the second is the acquisition of cell polarity and the generation of an epithelial sheet. A third element is cell-cell communication, a property inherent to signaling pathways. Once cells are in contact and adhering, they are required as a result of the two previous elements, to form an adaptive organism. Given the universality of the gastrulation process, it is assumed that such formation of a "tube within a tube" affords an adaptive capture of single celled prey within a primitive "gut" or digestive zone. The capture of single-celled organisms increases with the length of the digestive tract, and mobility is also thereby facilitated. Gastrulation is then required to follow seamlessly or naturally from the combination of cell-cell adhesion and cell-cell communication.

Cells had apparently discovered cadherins, the mediators of adhesive connections before the Cambrian 'explosion'. The evolution of animals (metazoans) required genomic innovations that allowed cells to adhere and communicate, and also provide the structural basis for necessary developmental processes. Such vital processes include tissue morphogenesis, cell sorting, and cell polarization. However, cadherins are lacking from all other multicellulars, such as plants and fungi. In fact, cadherins have only been found in metazoans and their closest single-celled relatives, the choanoflagellates. The common ancestor of choanoflagellates and metazoans was most likely unicellular (Abedon and King, 2008); the absolute and relative cadherin gene numbers in choanoflagellates are comparable to those of diverse metazoan genomes despite the fact of the latter's much greater morphological complexity.

Cadherins are differentially expressed during development and in adult organisms. Many cell types express multiple cadherin subclasses simultaneously, with the combination differing with cell type. It is then inferred that the adhesive properties of individual cells are governed by varying the combination of cadherins. Altering the normal expression of cadherins would be a natural suspect in metastasis of tumor cells.

The connection between cadherins and β-catenin, part of the Wnt signalling pathway (Cox and Peifer, 2004; Nelson and Nusse, 2004; Gottardi and Gumbiner, 2004; Korswagen et al., 2000) is an important regulator of cadherin-mediated adhesion, linking cadherins to the actin cytoskeleton, and suggesting an important role for the Wnt signalling pathway at the origin of multicellulars, as well as an origin of diploblastic animals before that of the bilateria (Barker and Clevers, 2000). In metazoan epithelial cells, recruitment of β-catenin to the plasma membrane facilitates essential interactions besides adhesion, and especially interaction between cadherins and the actin cytoskeleton, the latter maintaining cell shape and polarity. The formation of epithelial sheets then becomes possible with the necessary combined action of β-catenin and cadherins (Nelson and Nusse, 2004). Apparently there are two conformational forms of β-catenin, one involved in the "canonical" Wnt pathway to the nucleus and subsequent transcription, the other form influencing adhesion by interaction with cadherin (Korswagen et al., 2000; Barker and Clevers, 2000 ).

The adhering polarized cells form a closed hollow spherical structure with thickness, a blastula. As cell division proceeds, the size or the total area of the blastula increases. Perhaps energy minimization confers a spherical shape to the hollow thick walled blastula before the onset of shape-changing patterning occurs. One of the requirements of any plausible model is that it describe the invagination of this hollow epithelial sphere, and only after it reaches some specified critical radius, or equivalent total area. This critical radius should be given in terms of the parameters of the model. A second and even more basic requirement of any patterning model is that it provide, at each developmental cycle, a specified positional 'niche' for stem cells. Presumably stem cells are not positioned at random in a tissue. A speculative possibility is that stem cells occur in the 'Margin cell' region, but much more is needed to specify an actual stem cell. The present model does not specify such mechanisms.

One virtue of the basic patterning model discussed below, and in Appendix A, beyond its simplicity, is that is has a plausible interpretation in terms of signaling pathways, or rather, in terms of the interaction of two signaling pathways. An activated signaling pathway results in activation, via the action of various intermediate factors, of a specific transcription factor in the cell nucleus that switches genes (or gene networks, on or off. It is sensible, then, to ask whether such can provide basic patterning, and if so, how.

Which specific two pathways are active are a (possible) variable at each cycle. A cycle changes upon the switching from one signaling pair to another. As is argued here, the interaction of two signaling pathways provides pattern formation due to the most plausible of interactions. There is no need to discuss whether or not the "morphogen" travels around or through the cells, nor how far such morphogens travel (Gurdon and Bourillot, 2001). No long range diffusion is involved in the patterning formation in the present model; the ligands do not travel but one or two cell diameters before being captured by their respective receptors. Ligands are highly attracted to their respective receptors (Kerszberg and Wolpert, 1998).

There have been many suggestions for the origins of pattern formation, too numerous to review here (e.g., Murray, 1990; Koch and Meinhardt, 1994; Pilot and Lecuit, 2004; Teleman, Strigini and Cohen, 2001). One common model of patterning proposes that a morphogen gradient is established by a diffusing morphogen(s) emanating from a source. Arguments against morphogen movement by long-range diffusion have been raised by many, including Kerszberg and Wolpert (1998), who asserted that capture of morphogens by receptors so impedes diffusion that useful stable gradients are unlikely to arise by that mechanism. They proposed that morphogens instead use a sort of relay mechanism in which receptor-bound morphogen on one cell moves by being handed off to receptors on an adjacent cell. A more recent review by Kerszberg and Wolpert (Kerszberg and Wolpert, 2007) raises several further difficult problems with the orthodox picture of pattern formation.

Here a quite different picture of pattern formation is proposed. The only diffusion involved is short range, involving the diffusion of ligands over at most several cell diameters, or even between nearest-neighbor cells. Since the term "morphogen" has come to be so firmly associated with long range diffusion, (in spite of its Greek root, as "shape-genesis") a new term is adopted here to avoid confusion. At each growth cycle, two pattern regions will be specified, and these two provide two different messages to the genome in the two non-overlapping space regions. As the spatial region grows, in general two different 'pattern regions' will arise from zero amplitude to cover the region, when the area of the (middle) epithelial sheet is adequate to accommodate the new pattern. As the latest pattern arises from zero amplitude, the two previous pattern regions subside in amplitude. At each cycle, two new lengths will generally be specified by the parameters of the model. A binary transcriptional mode of interaction with the genetic network is then provided in time, and a unique spatial specification is provided at each cycle. Each pattern region, of which there are two different ones at each cycle, describes a region which varies in the amplitude of signaling activation, and thus possible transcriptional effectiveness. Such gradient of signaling activation inherent in the model is necessary to provide directional information in the plane of the epithelium. The gradients may be supplied by the various factors associated with the particular Asp, one of which is often β-catenin.

The term "**Asp**" ('activated signaling pathway') will be adopted here to denote the density of activation of a particular pathway, and this will be taken to be also the density of ligand bound receptors of a particular type, i.e., signaling pathway. This will be normalized to unity, so that denoting a particular density by R, we have that $0 \leq R \leq 1$. (The activated receptor densities will be divided by a maximum allowed value). The pathway amplitude will be assumed proportional to the density of a specific factor, (for example, in the case of Wnt, the factor β-catenin), a factor that will result in the nuclear activation of a transcription factor of a specific type.

To give a more specific example, suppose that Wnt and Hh are the two signaling pathways at a specific cycle; their amplitude of activation will occupy two different spatial regions on the epithelial sheet, with only a small region of very small amplitude of each pathway separating the two 'Asp' signaling regions. Within each of the two regions of signaling activation there is in general a variation of amplitude; the β-catenin which grows in cytoplasmic amplitude upon activation of the Wnt pathway will vary in amplitude in cells across the region which will contain numerous cells in general, as will also the density of ligand- bound-receptor density of this pathway. Thus gradients will exist in the plane of the epithelial sheet; such gradients are necessary, for only one example, to give a direction to hairs on fly legs, etc. (Kerszberg and Wolpert, 2007).

The present model is based on the concept that freely diffusing, non directional ligands in the extracellular space site onto specific cell receptors, which then activates a signalling pathway to the nucleus. The cells are thereby stimulated to emit in a concentration dependent way new ligand of the same type into the extracellular space, the amount emitted proportional to the level of excitation of the signaling pathway. The emitted ligand has very possibly been stored beforehand in vesicles in the cell at or near the cell membrane. The diffusion is short range, as the ligands have high affinity for their receptors. .

The present model most basically envisions two different 'Asps', along with a propensity (as an equivalent formulation shows) of these two Asps to avoid each other. The Asps are taken to be the density (number/area) of receptor plus its attached ligand in a small region, or more generally, an Asp is the density of a given activated pathway, leading to activation, through a series of intermediate non-nuclear factors, of transcription factor of specific type,. Each Asp has a threshold for activity, assumed the same for each Asp for simplicity, and our focus is on a desired steady state configuration for each total area. "**Asp**" is then used here to indicate only the level of excitation of a given signaling pathway, and also indicating the level of excitation of a specific factor (e.g., β-catenin) that will lead to activation of a specific transcription factor in the nucleus. There is then no discussion of whether "morphogen" moves passively around or through cells; it is actively secreted.

The explicitly time dependent pattern model involves four variables: two ligands, and two active receptors (Cummings, 2004; 2005; 2006). The **key elements of the model** can be stated most simply. Activation of one signaling pathway acts to deactivate production of ligand of the second pathway, while at the same time stimulating production of ligand of similar type. Differential equations for these four quantities may be written immediately, as shown in Appendix A, (also Cummings, 2004, 2006).

Regardless of details of the nonlinearities, and addition of other complications which may be added in an obvious way, several properties emerge to distinguish this tripartite model from others. 1) There is spontaneous Asp activation into two distinct regions at each cycle. Two lengths determined by the parameters of the model dictate the size of these regions. 2) The activation occurs from an original zero level of Asp density and does not depend on the presence of nonlinearities. 3) The model is directly motivated by the known ubiquitous involvement of signaling pathways in earliest embryonic development. Surprisingly, only a handful of pathways are involved in embryogenesis, and these are employed repeatedly in different contexts. The model requires them to be coupled sequentially two at a time. The signaling pathways are involved in gene network activation and subsequent protein production, a process considered to be a slower process than the establishment of pattern. Then a "pre-pattern" is provided, followed by cellular differentiation. 4) The relative sizes of two diffusion constants entering the model for the ligand short range diffusion are not constrained, e.g., patterning is achieved even when the two diffusion constants are equal, and the ligands are not required to travel more than a cell diameter or two. The diffusion of the free ligands is assumed to be usually of short range, although this is not a requirement of the model. 5) 'Niches' between the two determined regions occur. These two regions surrounding **Margin cells** may be determined by 'selector genes' (Gerhart and Kirschner, 1997).

These interstitial '**Margin cell**' regions are relatively small regions and specified by some threshold, and they separate regions of activation of factors of either of the two signaling pathways alone. The center of the (thin) Margin region(s) occurs where the two Asp regions have the same concentration ($R_1 = R_2$), specifying a "Margin line". This region may be affected in some unknown way by both of the transcription factors at once, or perhaps not determined at all. This region is indicated in the left panel of Figure 1 by the symbol 'S', and its role must remain mysterious at present, subject only to unavoidable speculation. Here in the 'Margin region', cell determination does not occur as in the regions where there is unique activation of either one pathway or the other. One possible, but very tentative and speculative interpretation of this region is that it is the region of patterning leading to activation of genes which are typical of stem cells, (such as, e.g., Sox and Oct). In favor of such an interpretation is that the '**Margin cell**' region will always be surrounded by active signaling pathways, which will act to regulate the stem cell-like 'niches'. Mutations affecting the surrounding signaling regions would be expected to have serious negative consequences for the developing organism in the case that this region can be in fact associated with actual stem cells. But such speculation is motivated by the need to find a pattern region for the actual ubiquitous stem cells.

A basic requirement of any pattern model is that 'niches' be specified. Stem cells are not distributed in the tissue at random, but presumably have a specific spatial distribution. Of much interest is the region between two Asp densities (e.g., Wnt and Hedgehog ('Hh') ) occurring in the present model. This is a region where both aps are acting together to determine cell fate. Presumably such fate in the Margin region is different from that of the two distinct regions where only one Asp is effective. This is shown as shaded in the gastrulation example of Figure 1 (left panel). However, it must be stressed that stem cell designation is much more complex than can be specified here, and this present speculation is made in a "where else" sort of desperation. It may be supposed that this Margin region is a region where neither Asps $R_1$ or $R_2$ is able to effect cellular determination as one or the other can acting alone. Since cells in this region are not to be determined by either Asp density (e.g., Wnt or Hh) acting alone. It seems reasonable that they be designated as '**Margin cells**' at present, for lack of a better designation. This designation will minimally serve as a reminder that actual stem cells require a designated pattern position, and this region is the default option.

Typical stem cell genes will be activated (e.g., Sox, Oct) in an actual stem cell region. The Margin region is denoted in Figure 1 (left) as "S", and in Figure 1 (right) the corresponding region occurs near the blastula lip. Patterning of stem cells are expected to be of importance. Control of the stem cell proliferation then would fall to the cooperative effort of the two adjacent Asps, such joint control assuring that stem cell proliferation is toward definite adaptive ends of the organism. Aberrant somatic stem cells are probably the locus of tumor initiation, and Wnt and Hh pathways are known to function in the normal regulation of stem cell number in at least some tissues. Expansion of the somatic stem cell population may be the first step in the formation of at least some types of cancer (Taipale J., and P. Beachy, 2001; Reya and Clevers, 2005). It is to be expected that if the signaling pathways are involved in such a crucial manner in cell determination and cell shape, as well as control of stem cell fate, that cancer will be the likely consequence of disruption of such fundamental processes. Observation that both Wnt and Hh are so often involved in tumor transformations motivates one to seek a fundamental patterning model involving these pathways, as well as others (Plinkus et al., 2008; De Celis, 2004; Borycki, 2004). Clearly such speculations based on the model regarding stem cell patterning are most tentative.

Since, as proposed here, the interaction of two signaling pathways is crucial to normal development, one might expect that mutated genes associated with certain diseases (e.g., cancer) might be most significantly categorized by association with one of a handful of the most prominent signaling pathways of early development, rather than categorized by a much larger listing by gene specification alone.

The pattern sharpens as the amplitude of the two determined **Asp** regions increase with cell division and growth, and as nonlinear terms come into the picture. A possible way in which cells from two distinct determined regions (say, Wnt and Hh) can become separated by a single cell diameter is for the epithelium to bend in shape in the Margin cell region, and ingress into the interior, while bringing the two determined regions into close contact. The ingressing cells may be expected to contribute to the mesoderm (Figure 3a).

To date there has not been an empirical characterization of the stimulation of further ligand production as the result of activation of a receptor by its target ligand, as presently proposed. This must serve as one prediction of the present model. The usual description of signaling pathways leaves the origin of the ligand that activates a given receptor unknown, except to characterize it as "emitted" and subsequently "captured" by its receptor. The 'why' and 'where' of the emitted ligand are usually left unspecified. Presumably such emission is not at random times, or at random places; if it takes place only from specific cells, a model specifying such cells or cell regions (e.g., "organizers") is required. ("Sources" and "sinks" must also be positioned). Here we reasonably assume that the amount of emitted ligand from a given cell increases as receptor activation by its specific ligand increases; activated receptor of type "$R_1$" stimulates ligand emission of type $L_1$ in the same small region. (It is interesting to ask how this could be otherwise). Such emission may be occurring by a "non-canonical" pathway, (perhaps where vesicle loaded ligand is already present), i.e., one bypassing the nucleus, although this issue is left unspecified by the model.

A very common view is that cells passively receive cues from a protein ('morphogen') passing around them, but this has been challenged (Ramirez-Weber and Kornberg, 1999; Kornberg and Guha, 2007). Rather it seems that cells play an active role in determining patterning (see Pearson, 2001). The uptake and subsequent release of ligand by individual cells appears to shape the pattern. In fact ligands have been seen in fruitfly wing discs in small round dots inside cells, in membrane bound vesicles (endosomes) that form when substances are actively taken up by cells. A speculation is that the stimulated emission postulated by the model may be from such stored ligand (or their precursors) in vesicles, prepackaged ready for transfer across the plasma membrane upon stimulation. Such vesicle storage may even presage the stimulated emission from vesicles occurring in neurons, very elongated cells; at any rate, there appears to be a rather rapid evolution of neuronal cells at about the Cambrian.

The second key requirement of the model is that as receptor activation of one kind increases, (and from what was said just before, 'like' ligand emission from that cell correspondingly increases), the emission of ligand emission of the second pathway equivalently is caused to decrease. Then, activation of one pathway acts to switch off or decrease activation of the second pathway, and vice-versa. This may happen in a number of ways, not yet determined, and such 'switching off' of one pathway by activation of another would be supposed to be an unchallenging affair via action of transcription factors. The simplest scenario to imagine, however, is where the first ligand is simply captured by the second receptor, such capture then acting to block activation of the second pathway.

The simplest consequence of the model of coupled pattern and geometry in axial symmetry is developed in what follows, namely gastrulation. The pattern model is always necessarily coupled to shape changes in the thick epithelial sheet. Gastrulation here means the formation of a tube within a tube, with the ectoderm differentiated from the endoderm. As the 'animal' grows further beyond the gastrulation phase, the patterning mechanism partitions the epithelial sheet into ever more complex regions. Each cycle divides the available spatial region into two new signaling patterns, each Asp separated by the accompanying smaller Margin region.

By a 'cycle' is meant here the growth of the specified signalling pair from zero density up to some maximum, followed by the subsequent decay as growth occurs. The simplest model for the action of the resultant two different regions of signaling activation on cell shape suffices to give gastrulation; one region causes the local apical cell area to exceed the basal area, and the other pattern region causes the opposite cell deformity (basal area greater than apical) at that point (or rather, small region). One assumes that different genes are involved in these cell shape change.

Our starting point is the blastula stage, a hollow sphere of polarized epithelial cells. The blastula will begin the process of gastrulation at some point as cell division and growth proceeds, providing an ectoderm and endoderm (Cummings, 2004; 2005; 2006).

The 'Wnt' pathway is apparently most important at the very beginning of multicellulars (Kusserow et al., 2005). So far, no Wnt genes have been described in unicellular eukaryotes, or from cellular slime moulds, or from choanoflagellates. It may be presumed that the appearance of Wnt genes and the Wnt pathway itself was linked to the origin and evolution of multicellular animals from a single-celled ancestor (Taipale and Beachy, 2001; Cummings, 2006). The relatively rapid evolution of an original Wnt pathway into similar versions, involving different transcription factors, is surmised to have occurred about the time of interest, the Pre-Cambrian. It has been proposed that the Wnt and Hedgehog (Hh) signaling pathways are evolutionarily related. (Nusse, 2003). These two are the most likely candidates for the original two pathways whose interaction gave rise to the original multicellular.

The math model of Appendix A is the time-independent version of the model which will occupy us here. This consists of two (dimensionless) "Asp densities", say $R_1$ and $R_2$. These two vary with two coordinates, (u, v), which are always taken to be in the epithelial (middle) surface. The shape of the closed surface changes with cell division and growth. Growth is assumed to occur via the changing total area. The total area is a parameter of the model put in "by hand' as an increasing function of time, A(t). The pattern in the steady state regime (corresponding to a specific total area) involves only two main parameters, namely, the size of each Asp region. As the total size (total area) of the animal increases, parameterized by the total area 'A' of the animal, the maximum amplitudes of the Asp densities also increase, each in its specific region, up to a maximum. Two separate regions are marked out, always with a smaller **Margin** region separating the two. The small interstitial region between the two determined will be designated, as mentioned, as a 'Margin' region, whose genetic identity is presently unknown. It will be proposed that the intersection of two of them provides the basis for required precise patterning of the positions of eyes, antennae, legs, wings and gills.

Asp maxima density may be determined, one may suppose, by regulatory genes and switches. However, an amplitude cutoff is included as an integral part of the more complete nonlinear model. This maximum cutoff simply insures that the Asp densities stop increasing beyond a point, as the total area increases..

After achieving maximum amplitude, a given pattern must decay, and further growth again takes place. A new signaling pair is presumably specified by transcription factors activated in the previous cycle. When a total size is reached which can support the next, more complex pattern, this next pattern will begin to emerge, again rising from zero level of each of two (new) Asps. The next two Asps (defined as the densities of activated signaling pathways) may be the same signaling pathways as the previous two, or arise from the coupling of two different pathways. The general rule is then that a pattern first forms when the area is sufficient, as dictated by the two length parameters involved. Which pattern of several possibilities will begin to rise from small amplitude will in each pattern cycle be determined by the geometry of the region, and also by conditions on the 'boundary' of the region, and importantly by transcription factors determined by the previous cycle. This implies that coupling of pattern to geometry is a crucial element in the discussion to follow. The pattern (the 'Asps') directs geometry, most likely via the genes specified by the transcription factors specific to a given signaling pathway, and the epithelial sheet geometry in turn directs patterning, as growth is ongoing (Cummings, 2004, 2005, 2006).

In each successive pattern cycle, pattern growth followed by decay, gene network activation is determined in a binary fashion, giving possibly unique and different determination in each of the two regions specified by their Asp densities, and excluding the "in-between" Margin cell regions. Each cycle of cell determination overlays the previous.

The simplest result of the model is that no pattern is allowed on a blastula whose radius is below a certain value, as determined by the model length parameters in the linear regime. This is a minimal requirement of any model, namely that a onset blastula radius be specified, that is, the radius just before gastrulation begins. As the sphere radius of the blastula grows, at a certain critical radius a pattern, of initially small amplitude, begins to emerge on the blastula beyond this specific radius. (This condition (Appendix A) is that the squared radius $R_0^2$ times $(k_1^2 + k_2^2)/2$ must be equal to unity; here the two k's are inverse lengths defined in terms of the model, and are the only two of the three parameters of the pattern model). The pattern is such that Asp $R_1$ begins to emerge in the upper hemisphere (say), and the Asp $R_2$ emerges in the lower hemisphere, with each of the two maximum amplitude at the opposite 'poles'. As these amplitudes grow, for instance, Asp $R_2$ causes cell shape changes in the lower hemisphere, such shape changes becoming larger as the amplitude of the Asp increases with total area increase. The epithelial cells in the lower hemisphere change initially so that the basal cell surfaces increase relative to the apical surfaces, with the opposite happening to cells of the upper hemisphere, and with perhaps smaller amplitude. A 'gastrulation' process then occurs. As the overall animal surface area increases, so also do Asp amplitudes, and the sharpness of delineation of the two Asps densities increases, due to the increasing effect of ever-present nonlinearities.

Such a coupled, changing geometry-Asp model has been carried out numerically for this simplest situation of axially-symmetric gastrulation for numerous total areas (Cummings, 2004, 2005, 2006). The shape of the gastrula depends as expected on the specific parameters assumed, and on the dependence on the Asps assumed for the curvature (Cummings, 2005, 2006); but invagination, "a tube within a tube" is the general robust result. Figure 1 illustrates two typical resultant 'gastrula' surfaces as well as the Asp densities $R_1$ and $R_2$, for two different total surface areas, and for a specific nonlinearity, the nonlinearity non critical, and simply limiting the growth of the Asps beyond a given density. The coupling of Asps to geometry accounts for the most severe nonlinearities, via the Gauss equation relating the Gauss curvature K to the Laplacian of the logarithm of the metric of the surface (Cummings, 2005, 2006). The signaling pathways can affect cell shape changes either by canonical pathway products of the activated receptors, via proteins that affect the (three) filament types that determine cell shape, or by a non-canonical Wnt pathway known to generate cytoskeletal changes as a consequence of RhoA activation (Povelones and Nusse, 2002).

Margin cells situated at the blastopore lip as occurs here may possibly have lowered adhesive affinities than the cells of the two determined regions of ectoderm and endoderm, and thus be able to migrate to the space between these two ectoderm and endoderm layers to begin the formation of mesodermal tissues such as heart, lungs, and neurons found in bilateral animals. Further, this invagination process may sensibly act to bring cells of two different determined region (e.g., Wnt and Hh) adjacent to each other, as may be required in, say, segment formation.

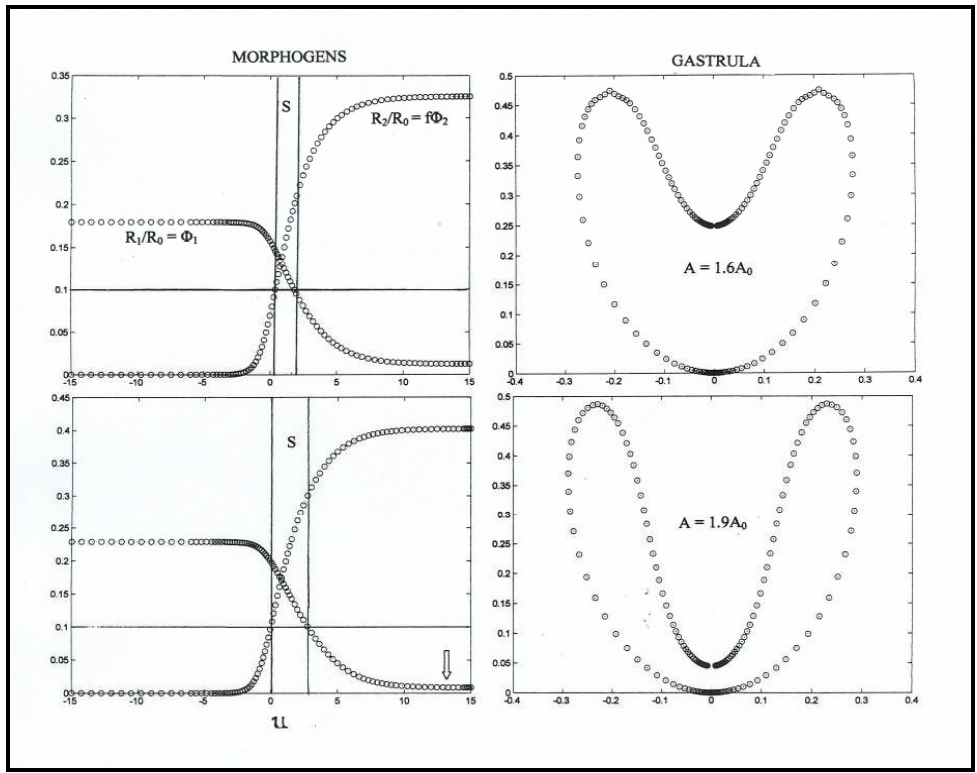

# Figure 1
**Figure 1:**  left: Asps;  right: Invaginations

The left column shows two numerically obtained 'Asp' solutions, each frame showing two interacting Asps, for two different total areas indicated. There are three coordinate 'u' regions shown for each total area, one (left-most) corresponding to Asp density $R_1/R_0 = \Phi_1$, and the right-most corresponding to the Asp density $R_2/R_0 = \Phi_2/f$. Between these two regions is a third region, for each total area A, between the coordinates $u_1$ and $u_2$, where the two Asps are under the influence of both Asp densities, and thus is a region of different determination of either one Asp alone. This is designated by 'S' as the 'Margin cell' region, and is located near the blastopore lip on the two corresponding axially symmetric shapes shown on the right hand side of Figure 1.

It is worth noting that the mathematical equations resulting from either the simplest "two-Asps-that-avoid-each-other" approach (Cummings and Strickland, 1998) or the present more biologically motivated interacting-signaling pathway approach, both lead to, in the linear version, and when not coupled to geometry, two well studied equations of mathematical physics. These are the Helmholtz equation of wave theory, and the even better known Laplace equation, the latter occurring in a number of disciplines, such as (e.g.), fluid flow, electricity and gravity. The appearance of these basic physics equations as a result of patterning by the interaction of signaling pathways may provide new insight into the famous work of D'Arcy Thompson (Thompson, 1942). In Thompson's work, the laws of physics were frequently taken to be at the heart of many observed natural patterns. Here, rather, similar patterns result from the interaction of the signaling pathways, with resultant cell shape changes. Biological "laws" may then bear strong resemblance to laws of physics.

## 3. On the divergence of Urcnidaria and Urbilateria

The main groups of bilateral animals are deuterostomes and protostomes. Their last common ancestor is called Urbilateria (Robertis and Sasai, 1996). It is interesting to remember that certain features of animal body plans have been conserved since the Pre-Cambrian (Hobmeyer et al., 2000; Finnerty, 2003), among the most notable being the bilateral body plan. Another feature thought to have been conserved from the Cambrian is the ubiquitous segmentation in annelids, arthropods and vertebrates. At the species level, on the other hand, many changes have accumulated. It is then of interest to ask if some of the important conserved features, such as bilaterality and segmentation, may result from strong biases at work in animal patterning. Rules of patterning then may act as another kind of "selection", acting along with but constraining natural selection. Such is the contention and focus of the present section, and also in the case of plant patterns in the following section.

Cnidarians (such as sea anemones, hydra and corals) and Poriferans (sponges) split off before bilaterians. Their last common ancestor may be dubbed "Urcnidaria". The discussion from this point continues where Figure 1 leaves off, at the point where the endoderm is shown as about to come into contact with the ectoderm (in the bottom right panel of Figure 1). Concern then centers on the new boundary conditions set up when contact occurs between epithelial surfaces. At the point of contact of endoderm and ectoderm along the axis of Figure 1 there are two very different scenarios to be considered. In the first, the surfaces come into contact, new boundary conditions are set up, but the two surfaces do not interpenetrate. Cnidaria-like animals result in this case.

The second scenario and the one of focus in this section, is a situation when the surfaces come into contact at the animal and vegetal poles, cell death occurs in the region of contact, the surface reforms and a new opening is produced at this contact point or region. In the first case of no new opening, the surface remains topologically equivalent to a sphere. In the second case of a new opening, the topology is abruptly changed from a topological sphere to a topological torus, or 'donut'. These two scenarios correspond to the genesis of the two sister groups, cnidaria and bilateria, now referred to as "Urcnidaria" and "Urbilateria". We investigate the linear model of Appendix A, inquiring for directions based on the model for the subsequent evolution, starting now from the crucial point of contact. Growth subsequent to the contact is such as to produce an ever increasing elongated, hollow donut structure. The geometry is simple, namely the Gauss curvature K is zero, and the metric 'g' entering the Laplacian is constant.

It is not the intent here to discuss the cnidaria patterning based on the model in any detail. Cnidaria typically have a third layer of (mostly) non-cellular mesoglia between the two distinct tissue layers. These diploblastic animals have a mouth, which doubles as an anus, and typically a ring of tentacles around the mouth. As boundary condition, it is reasonably supposed that the two Asp densities at the point (or rather small region) of contact are such that the two Asp densities are equal to the same constant, so that $\Phi_1 = \Phi_2$, and these values now become fixed there. A cnidaria-like animal is the result of the touching boundary condition (Cummings, 2006), only in the sense that a periodicity around the mouth region is accompanied by an exponential decrease of each **Asp** down the body axis, and alternating around the body axis. The periodicity in the radial direction is proportional to the mouth radius, as is found in hydra (Bode and Bode, 1980). The smallest radius would correspond to a bilaterally symmetric animal, as two distinct selector gene regions would predict (Finnerty, 2003). At the same time that the radius increases it is accompanied by a corresponding increase of the length of the body column; when a bud(s) can eventually form for a large enough area, and great enough length, as occurs in hydra. The model becomes vague at this point.

Interest focuses instead on the patterning of the 'Urbilateria' according to the model. In this case there is a transformation from the topology of a sphere to that of a torus or 'donut'. Based on the model, it is argued that the double periodic boundary conditions lead naturally to the development of bilaterally segmented animals, as one possible result. There are several possibilities, and precursors to segmentation are most apparent and ubiquitous in arthropods, annelids, and vertebrates. Here one interest is in the earliest development of these particular properties, bilaterality and segmentation. Insight into the origin and evolution of segmentation is central to understanding the body plan of the more these advanced major animal groups, arthropods, annelids, and vertebrates (De Robertis, 1997).

Instead of attempting to solve a very involved numerical model when surface shape is coupled to Asps (pattern) and vice versa, (as was carried out in the axially-symmetric gastrulation case, and resulted in the shapes shown in Figure 1 leading up to the point of animal-vegetal contact), an approximate approach is taken subsequent to the contact and formation of the second opening and "through-gut". Three approximations are involved. 1) The geometry is assumed to be cylindrical, described by only two parameters, cylinder height (or length) 'L', and radius 'R'; 2) The thickness of the epithelial sheets, endoderm and ectoderm, are ignored; and 3) the linear model is used. In spite of such apparently drastic simplifying assumptions, it is believed that interesting key aspects emerge, aspects reasonably expected to remain in evidence in a more complex or complete model.

The linear version of the model will, strictly speaking, be applicable only in the small amplitude regime. However, the linear version will indicate the general form of the pattern, even after it has risen in amplitude. This approach is adopted largely because the numerical simulation of a more extended model, with geometry coupled to Asp in two dimensions, is very much more involved numerically. Numerous general points are able to be made sufficiently convincingly with the simpler linear version, points reasonably expected to be carried over in the nonlinear model.

Two coordinates are pictured as always residing in the epithelial sheet, in particular, in the apical-basal bisecting ('middle') surface. It is worth remembering that the coordinates are always purely imaginary constructs of the observer, and consequently may be drawn arbitrarily. The pattern, on the other hand, is the 'property' of the animal. Any generally valid formulation must then be **invariant**, that is, independent of coordinate choice. The Laplace operator is a known invariant, and the Asps (the R's) are also invariant (scalar) functions.

It may be wished that the coordinates are chosen cleverly or conveniently in order to make the discussion as transparent as possible. Cylindrical coordinates are then used in this simple geometry, with 'z' describing length along the cylinder (inner and outer surfaces, in the case of a 'donut') axis, while an angle 'φ' describes a point at any given value of z. Figure 2a shows a cylinder of length 2L and radius R, and closed at both ends by circular discs. It is imagined, as shown in Figure 2b that the top half of the rather 'elastic' cylinder of Figure 2a is 'tucked' into the bottom half, forming an inner and outer surface. Both inner and outer surfaces are imagined to be of negligible thicknesses. A caricature of the shapes is produced in Figure 2b. The Figure 2b indicates a shape for which both 'disc' ends have disappeared, but the sheets have been joined at $z = 0$ and $z = \pm L$ to form a topological torus or 'donut'. It is this geometry on which we solve the linear model for the Urbilateria. Cylindrical coordinates with $-L \leq z \leq +L$, and $0 < \varphi \leq 2\pi$ describes length along the cylindrical axis, and an angle around the cylinder.

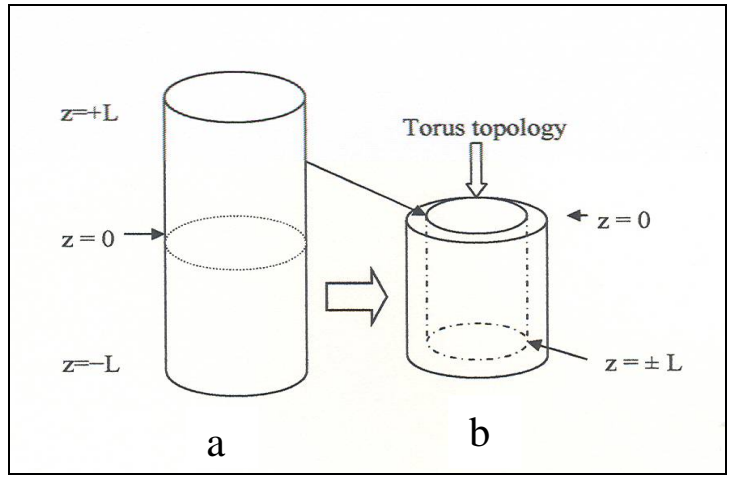

**Figure 2**

**Figure 2:** Cylinder Geometry for Urbilateria
A single walled cylinder of length 2L is shown in Figure 2a, and the corresponding double-walled cylinder is shown in Figure 2b. Figure 2b is visualized as obtained from Figure 2a by moving the top part of the cylinder (of length L) into the lower part (also of length L), so as to form an analogue of a gastrula, with an ectoderm and endoderm. The original two flat-end discs which would have coincided at the bottom of Figure 2b are now removed, so that after joining the edges, a topological torus or 'donut' is achieved. The thickness of the two inner and outer sheets indicated in the Figure 2b has been ignored in the analysis.

The solution of the model equations with the appropriate boundary condition may now easily be obtained analytically, and is given in Appendix B. The analysis hopes only to give insight into the potential of the patterning model. The boundary condition is that the pattern be doubly periodic, that is, periodic both along the body axis, (in the 'z' direction) as well as around the axis (the angular coordinate 'φ'). The geometry may be visualized by imagining a (hollow) 'donut' stretched appropriately along its axis. It is to be remembered that the endoderm has previously been differentiated relative to the ectoderm during gastrulation and before contact occurred. The solutions shown in Figure 3 are then only the new 'next' pattern superposed on the previous.

The relation required by the model (Appendix B) between radius R and length L is simply that

$$1 = (\pi m/kL)^2 + (n/kR)^2. \qquad (3.1)$$

This is the geometry at the point when the new (next) 'Asp' pattern just begins to emerge. The integer 'm' arises from the periodicity in the 'z' direction along the 'donut' axis, while the integer 'n' is associated with periodicity in the angular variable 'φ'.

It is clear that eq. (3.1) allows a variety of solutions, beyond those involving segments. If m=1 and n=0, then kL=π; this is the solution just at the point of growth when the gastrula has achieved a second opening. Among the possible scenarios representing further growth is one in which kL>>π for m=1, n=1 and kR≈1; this may represent an un-segmented, bilaterally symmetric (n=1), through-gut animal, such as a primitive version of a nematode, or acanthocephala (a spiny-headed worm). Another result may envision an originally bilateral animal (m=1, n=1) later developing (by metamorphosis) into a radially symmetric animal, with m=1, kL=$2^{1/2}$π, as kR increased to kR=5·($2^{1/2}$), with n=5 (a proto-starfish).

Our interest now will be in examining the very adaptive case of kL ≈ kR initially, and with growth, giving kL >> kR. This case is termed adaptive because it facilitates locomotion and also provides a long digestive gut. Whether bilaterality (n=1) develops at about the same time, (or possibly even earlier), as the second opening beyond development of the blastopore, or rather, bilaterality develops somewhat later, is left open. The assumption is that some animals may develop one way and other animals the other. The mesodermal tissue will always come from the existing ingressing Margin cells, in either case. Growth always occurs at the posterior Margin region.

Imagine a growing embryo, growing initially along the axial direction, and that initially kR < 1, and shown in Figure 3a. Figures 3a, b and c each show the cylinder patterns in schematic side view. A 'proto segment' (called a 'segment' here) consists of one growth cycle, and is signified by two different colors in the Figures. Each growth cycle activates different factors in each of the two sectors of the segment. These transcription factors will activate a different cascade of genes at each cycle, for example, besides, e.g., Wnt and Hh represented by the two colors of a segment. Hox genes may also be activated, differing from segment to segment. The vertical lines represent Margin line boundaries between segments, while the horizontal lines also represent Margin cell regions, or Margin 'lines'; all vertical lines are actually circles around the cylinder. The intersections of two Margin lines are specific points of interest.

The integer 'm' will first be '1', then '2', … as axial growth occurs, while n = 0; an axially symmetric structure is described. It may also happen that n=1 and bilaterality develops anywhere in time along the axial growth if radius permits. The case of L growing first, and n = 1 developing bilaterality later is shown in Figure 4. Growth occurs until kL = $m_0$π, for some integer m = $m_0$, while n remains zero, n = 0. The radius kR is too small for an axial pattern to appear at this stage. If m=1 when n=1, then bilaterality develops at the end of the gastrula phase, upon development of the second opening and a through gut.

The term "segment" here denotes the periodicity made of two adjacent 'Asp' regions (Gerhart and Kirschner, 1997). The Margin cell regions are regulated by the surrounding (and determined) 'Asp' regions, for example Wnt and Hh. Axial growth occurs in Figures 3 and 4 from the Margin cell region separating the posterior (light blue in Figure 3) segment and the adjacent light green segment. Such animals ("short-germ" arthropods) only grow from their posterior regions (Jacobs, 2005). It is known that Hox genes are arranged anterior to posterior in the same sequence as on the chromosome. Hox genes may distinguish segments between the anterior and posterior ends of the animal, affecting, e.g., the placement and number of appendages. The Margin cell regions (shown only as thin lines in Figures 3 and 4) bisecting those segments (except the growth region) supposedly represents tissue that migrates inward to the space between ectoderm and endoderm in Figures 3 and 4 to become mesodermal tissue, for example, muscle and nervous system.

The anterior and posterior will also be determined by the two different Asps, shown in Figure 3 as deep blue at the anterior end, and light blue at the posterior. The anterior and posterior are differentiated from the beginning of axial growth, as shown in Figure 3a, when there is only one anterior segment and one posterior determined region. The anterior ('A') and posterior ('P') remain differentiated by Hox genes from the abdominal or thorax segments as growth proceeds.

Upon growth in the radial direction, so that kR increases to the extent that kR = 1, n must remain zero; n cannot equal '1', since that would require m = 0 from eq. (3.1). So the integer 'n' will remain zero, and no new axial pattern develops until R increases to a value such that $kR = 2^{1/2}$. Then n becomes unity, and kL increases to ($2^{1/2}$ mπ). A new, bilaterally symmetric pattern then emerges. Signaling pathways pairs, such as Dpp and Hh protein, are presumed to be active at this point in bilateral determination.

Growth in the radial direction is shown in Figure 4. No Hox distinctions are shown, such as those possibly distinguishing thorax from abdominal segments. Figure 4a shows the new angular pattern; the length of each segment has increased by a factor $2^{1/2}$ in Figure 4 over Figure 3, depicting the axial growth which accompanies the angular growth. A bilateral animal has appeared, based on eq. 3.1). Presumably a new pair of signalling pathways becomes active at this point (e.g., BMP4 may be expressed ventrally, and Dpp expressed dorsally (Finnerty, 2003) as in the bilateral determination of the fly). Different pairs of pathways perhaps are expressed for each species, and different from the pair giving rise to the axial growth pattern. The next small amplitude pattern with m = $m_0$ = 2 and n = 1 begins to emerge, superseding or overlaying the previous m = $m_0$, n = 0 pattern. The value of n = 1 establishes bilateral symmetry, here initially with $m_0$ (here m = 2) segments, the angular pattern being indicated in Figure 4a. The two Asp amplitudes are schematically represented in Figure 4 by the 'solid' (colored) and 'empty' rectangles, respectively. Figure 4a also implies equality of two Asp concentrations at two angles around the cylinder, (one not shown, and occurring at opposite sides of the cylinder; remembering that the cylinders are all shown in side view of the Figures 3 and 4 means that only one of the two equal axial (horizontal) lines is shown, i.e., running down the center of the Figure 4a).

This new pattern indicated in Figure 4a overlays the previous pattern with n = 0. In each case of the decay of a pattern, and subsequent growth of a new pattern, binary genetic decisions will be made, according to the two Asps involved. Different colors have been used in the transition form Figure 3 to emphasize that in general two different Asps are involved in the two situations, each pattern involving different transcription factors. According to the model, a binary genetic tree is then generated, binary genetic decisions being made at each transition from one pattern to the next. In principle, a unique genetic specification for each region at each cycle is then specified.

As growth proceed, kR may be expected to increase further to kR ≈ $2^{3/2}$, and the integer n will increase to n = 2. The radius may stabilize at this value (or even grow further). It will be most adaptive for the length to grow very much relative to the radius, although Figures 3 and 4 only pictures m = 4, and four segments. There are now 16 points of intersection in all for m = 4 and n = 2 shown, with only 8 intersections shown in the side view of Figure 4b. Natural selection apparently often prefers a long digestive tract, with $kL \approx 2^{1/2}\pi \cdot m$, $m = m_0 \gg 1$, also providing for relatively rapid locomotion. Bilaterality in turn provides for directional, stereo sensors. A bilaterally symmetric, segmented 'animal' with a through gut is now on hand via the simple model.

Figure 4b illustrates the pattern corresponding to n = 2 and m = 4, also in side view. What is to be especially noted are the intersections of two lines shown along the length on one side of the cylinder with thin circular lines (shown as vertical in the figure) around the axis. These are Margin cell regions. There are two other horizontal lines on the opposite side not shown. These intersections represent small locations that correspond to special Margin cell regions. It is proposed that the two pairs of locations corresponding to each of segment, dorsal and ventral, provide natural loci for later appendage outgrowth and development, as directed by Hox, Dll, and Pax-6 among others, by way of regulatory genes and associated switches (Carroll, 2005; Prud'homme, Gompel and Carroll, 2007).

Pairs of points situated ventrally (the lower set of pairs in the Figure 4b) may designate location of limbs such as legs (Tabin, Carroll and Panganiban, 1999; Lecuit and Cohen, 1997). Hox genes may increasingly specialize segments along the A-P axis. Dorsal (horizontal) Margin cell intersections with vertical Margin cell regions (actually circular) of Figure 4b may designate gill positions, or arthropod wing loci. It is interesting to compare cross-sectional views of fruitfly development (Bier, 1999, p. 375) with the patterns of Figure 4a and Figure 4b. In comparison, the Figures 4a and Figure 4b must be considered as 'overlaid'.

The point to be emphasized here is that the position of legs, gills, wings, etc., must be specified rather precisely, by some patterning mechanism. The animal's eyes, antennae and legs are to be placed with precision. Cell-cell communication is essential in this respect, and such positioning cannot be accomplished by reference solely to genes turned on or off in a single cell, and not by reference to genes and switches alone. Figure 4b suggests a method for such positional specification, albeit illustrated in an oversimplified geometry; crossing Margin cell regions may position outgrowths such as limbs, gills and wings. The proximo-distal outgrowths specified by the distal-less (Dll) gene (Carroll, 2005) occur according to the model at positions specified by the intersection of two Margin cell regions. Growth occurs only from Margin cell regions, and the active distal-less (dll) gene is necessary to originate any outcropping from the main body. The almost exact spatial position of "master regulatory gene" action must be supplied in each case. The Cambrian "Hallucigenia" is an interesting early animal in this respect, where the position of seven legs and seven (presumable) gills are specified. The gene Pax-6 specifies where an eye, fly eye or mouse eye, etc., is to be located, exactly and not approximately; again, where the Pax-6 gene places the eye must be specified precisely by any positional information model. The present model provides a possibility for such accurate placement.

A segment width according to eq. (4.1) is $\Delta(kL) \approx 2^{1/2}\pi$, and the corresponding radius will be $kR \approx 2^{1/2}$ (or greater) for a bilateral animal. Then it is expected (as observed, or predicted) that width $\Delta L$ will invariably be less than the segment circumference $C = 2\pi R$, or $\Delta L < C$. There does not seem off hand to be a compelling reason to rule out larger segments on the basis of selection; for example, a segment width several times the circumference, with perhaps each segment supported by a number of leg pairs might well be imagined. But such do not appear in nature.

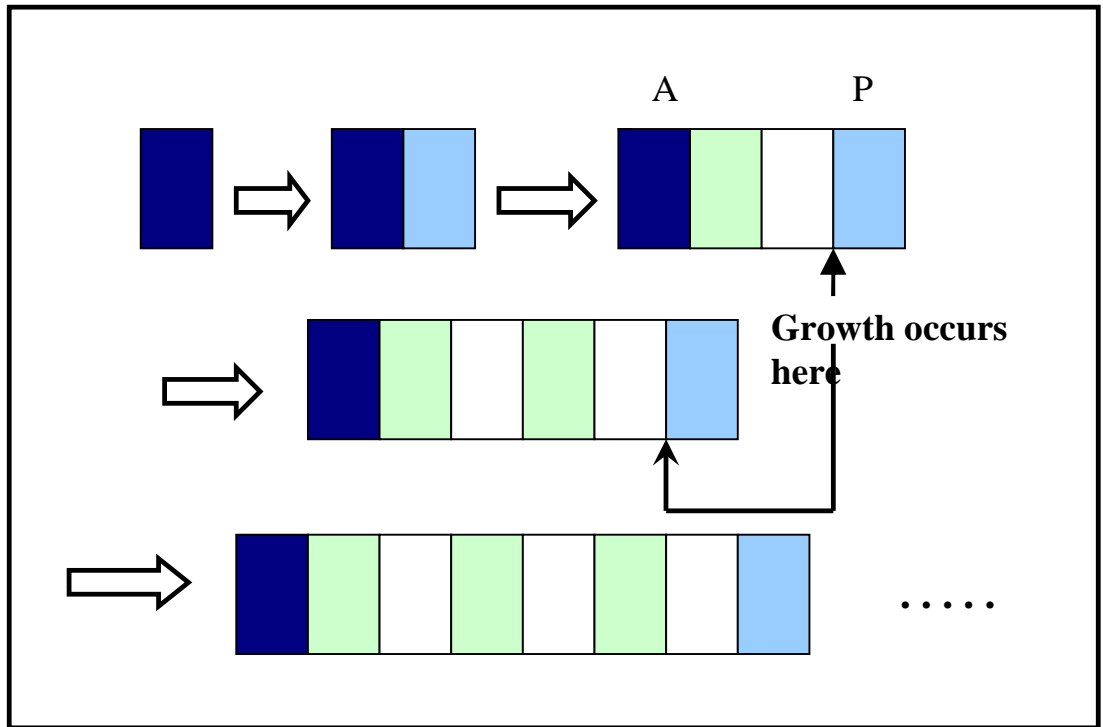

**Figure 3**

**Figure 3:** Patterns for axial growth
The open double-cylinders of 'donut' topology of Figure 2b represent the simplified geometry used to mimic the origin of the first protostome, or Urbilateria. As growth occurs, both L (length) and R (radius) increase, as indicated by the transition from Figures 3 to Figure 4. Segments are denoted by the combination of a light green region (an 'Asp') and a white region in Figure 3, with the green on the left side of a white region. Each picture shown in Figure 3 and Figure 4 are double cylinders seen in side view. Figure 3 (top) shows a schematic of the solution in which axial growth has occurred allowing one, two, and then three segments to appear. This organism has at this point axial symmetry.

Diverse bilateral taxa, including representative lophotorchozoa, ecdysozoa and deuterosomes, share aspects of a developmental process where repeated pattern elements are added posteriorly during development. This process of terminal addition suggests that it derived from a shared ancestral mode of development. Modifications of the process of terminal addition of repeated elements apparently occurred in the early Paleozoic radiation of Bilateria (De Robertis, 1997; Jacobs, Hughes and Winchell, 2005). Such terminal growth is consistent with the fact that the anterior/posterior sequence of Hox genes share the chromosomal Hox gene sequence.

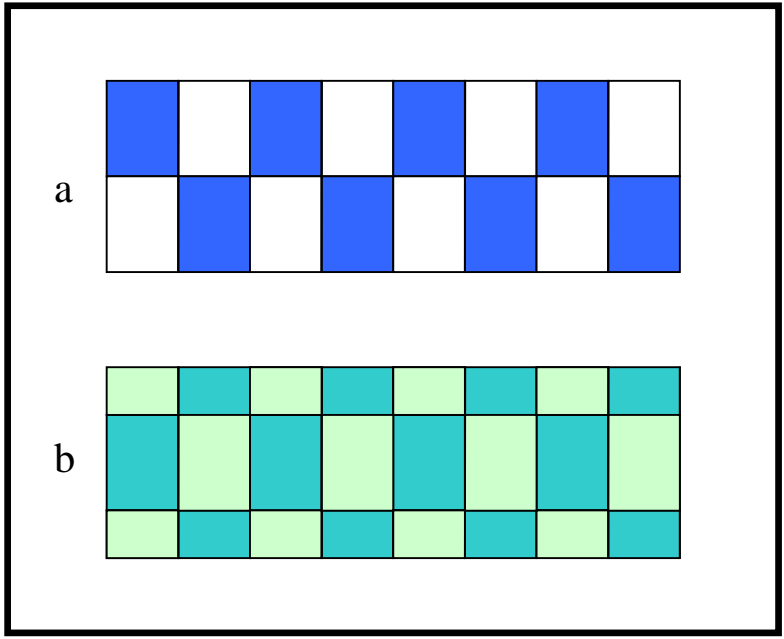

# Figure 4

**Figure 4** shows a next growth cycle beyond Figure 3, when the radius R has grown to reach a minimum radius of $kR = 2^{1/2}$, at which point integer n of eq. 3.1) is first allowed to become unity, signaling the origin of a bilaterally symmetric animal. The line running horizontally down the center of Figure 4a indicates the location where the Asp amplitudes (one blue and one white to emphasize the new Asp pair) are very small, thus specifying a Margin cell region as a thin horizontal line. There are four segments, and segment boundaries are shown by a heavy black line in Figures 4a and 4b. Further radial growth, as indicated in Figure 4b, starting at a minimum radius of $kR = 2^{3/2}$ (when m = 4, and n = 2 in eq. (3.1)), gives a situation where each segment is divided, (for each of endoderm and ectoderm), into **eight** regions. Filled (dark green) and light green indicate the regions of dominance of the two new Asps following those of Figure 4a. Figure 4b is an animal with 28 intersecting ectodermal Margin cell intersections or 'points'. If leg pair positions are designated by the ventral longitudinal line intersections with the circular Margin lines, i.e., thin Margin cell regions, an animal's leg pairs are precisely positioned.

      Interest turns to the growth and development of a typical "short-germ" arthropod. The growth of such an embryo on the basis of the present model is imagined to proceed as follows. At first there are two determined regions, denoting the two Asps in the first growth cycle; m = 1 in eq. (3.1), and shown in Figure 3 (top). This first 'segment' is considered as special, determined as different from the rest to follow, and distinguishing anterior ('A') from posterior ('P'). The segment is separated by a Margin cell region. One Hox cluster is present at this stage. Growth now occurs in the Margin cell region between the determined head and tail. In the next growth cycle, corresponding to m = 2 in eq. (3.1), there will be three regions separating the 'm' segments, with four 'Asp regions'. Growth is from the Margin cells adjoining the posterior region 'P' at each growth cycle. The Margin cells between the other Asp regions are then expected to divide and submerge (grow) into the interior, the region between endoderm and ectoderm, where they form mesodermal structures such as muscle and nervous tissue.

      Consider the possibility that limbs (or, e.g., lobopods) are generated at each ventral Margin cell **pair intersection** indicated in Figure 4b. For example, if a leg pair

(one leg on either side of the cylinder) occurs at each separation between Asp pairs in a segment, (e.g., Wnt and Hh or Hh and Dpp, etc.) then an animal with seven leg pairs will occur. Clearly only odd numbers of leg pairs can occur in any situation, for any length of axial growth; the Hox genes do not differentiate along the axial direction. Each growth Asp pair would have the same Hox genes active as far as leg development is concerned.

A **key point** is that such positional specification as provided in this way can give precise positions for (ventral) legs, (dorsal) gills and wings, remembering that Hox genes will generally act to distinguish segments in more evolutionarily complex animals such as the fly. It is further clear that the present pattern model provides specification of anterior-dorsal, anterior-ventral, posterior-dorsal and posterior-ventral specification for limb bud and future appendage. The gene Distaless (Dll) is a "master gene" thus activated, and providing proximo-distal instruction to the limb bud.

## 4. Plant Patterns

The patterns occurring on the stems of plants (Mitchison, 1972; Douady and Couder, 1991; Cummings and Strickland, 1998) is presented in light of the present model. The suggestion is that, even though the biochemicals involved in the two kingdoms, animal and plant, are expected to be quite different, the simple tripartite patterning model (two Asps and Margin cells) of Appendix A is assumed to have applicability to plants as well as animals. The aim is to show that while the present patterning model gives the known arrangements of plant patterns ("phyllotaxis"), it nevertheless explicitly forbids the number 'four' in a spiral arrangement with one 'leaf' on each level. This means that, counting up or down the stem from some initial leaf, (taken as at the origin) one leaf per level, the total number of 'leafs' until a repeat (a leaf directly above the initial leaf) occurs, 'four ' is forbidden by the present pattern algorithm. In fact, what is observed in nature is that the number four is strikingly less frequent than the numbers 2, 3, 5 or 8, which are Fibonacci numbers. This argues the case for bias in development, albeit now for plant rather than animal development. One is challenged to find a plausible argument from natural selection or adaptation for such paucity of 'four' as is observed to occur. There is clearly no structural reason that the spiral 'four' should not appear.

Outgrowth of 'leafs' from a common stalk again occurs from points of intersection of two Margin regions, as in animals. A prediction is that there will be found a "master regulatory" gene at these positions, as in the animal case. Such 'leaf' locations are shown in the contour map of Figure 6 as five circles, for the case of N=5.

The simple patterning mechanism of the present paper in fact reproduces all Fibonacci spiral patterns (Mitchison, 1972; Douady and Couder, 1991; Cummings and Strickland, 1998), as well as the common decussate and distichous patterns, and numerous others commonly seen, such as whorls. Decussate patterns are very common, and consist of two leafs at the same level, placed 180° apart on the stem, with the next pair up (or down) along the stem rotated related to this first pair by 90°. Decussate patterns are the simplest example of alternating 'whorl' patterns, which may have three, four, etc., leafs at the same level. Distichous phyllotaxis is also very common, where successive higher (or lower) single leafs are 180º from the preceding one, examples being corn, ginger and ferns. Superposed whorls also commonly occur, where, in the most common case, a pattern of two leafs situated 180º at the same level, followed at another level by two leafs directly above the original pair. This latter pattern is particularly common in compound leaves. All of these patterns are easily reproduced by the simple algorithm of Appendix A.

Appendix C gives solutions to the model as applied to plant patterns on a stem. A new pair of integers (p, q) is introduced by the model, which underlay and predict the more usual 'parastichy" integer pair (m, n) (Mitchison, 1972).

The unit square is bounded by the coordinates $x/x_o$ and $y/y_o$ in Figure 5. A positive integer pair (p, q), $p \geq q$, designates a particular pattern. The axial coordinate around the stem is $x/x_o$, so that all points at $x = 0$ and $x = x_o$ are the same. The unit length up (or down) the stem is taken to be $y/y_o$, and the line $y = 0$ has the same values as $y = y_o$. Interest is focused on a repeating pattern in the coordinate 'y' occurring at the point $y_o$, with $x/x_o = 0$ or 1. The solutions of eqs. (C.1) and (C.2) have been constructed so that there is always a zero of ($\Phi_1 = \Phi_2$) at the origin $x/x_o = 0$ and $y/y_o = 0$, (or the point (0,0)), as well as at the other three corners (1,0), (1,1), and (0,1). The repeating leaf at (0, 1) is not counted.

Intersecting lines of ($\Phi_1 = \Phi_2$) denote the positions of leafs or florets, and are located by requiring that the argument in each Cosine function of eqs. (C.1) and (C.2) be $\pi/2$ times an odd integer 'i' or 'j' in each case. This gives at once the two equations

$$y/y_o = (p/q)x/x_o + i/(2q), \qquad (4.1)$$
$$y/y_o = \pm(q/p)x/x_o + j/(2p). \qquad (4.2)$$

A given leaf is designated by a particular integer pair (i, j) designating an intersection of the two straight lines within the rectangle of Figure 5. A particular pattern is specified by an integer pair 'p, q'. Eqs. (4.1) and (4.2) are in a standard form. The two straight lines have slopes $S_1 = (p/q)$ and $S_2 = \pm(q/p)$, and the terms i/2q and j/2p are intercepts of the straight lines with the y axis. The y intercepts do not all lay within the unit square. Pattern construction follows easily from eqs. (4.1) and (4.2).The intersections of the lines shown in Figure 5 gives the positions of the 'leafs' in a given pattern specified by the pair (p, q). The first leaf is at the origin x=0, y=0, and is to be counted, while the repeat at x=0, y=$y_0$ is not.

The number of leafs in a given pattern, in the case that there is only one leaf per level of either x or y, and (p, q) are relatively primed, is given by the expression

$$N(p, q) = p^2 \pm q^2. \qquad (4.3)$$

This can be seen by eliminating x (or y) in eqs. (4.1) and (4.2) and observing that $0 < y/y_o \leq 1$, which implies that the maximum number of leafs in a pattern is given by eq. (4.3). In the case of the plus sign in eq. (4.3), the two sets of straight lines, corresponding to a set of integers 'i' in the one case and 'j' in the other, have opposite slopes, while for the negative sign in N the two sets of straight lines both have the same slope. This is shown in Figure 5 for the two cases of N = 3, N = 5 and N = 8.

When there are 'J' leafs on the same level, the expression for the number of leafs in a pattern becomes simply $N(p, q, J) = (p^2 \pm q^2)/J$. A 'whorl' pattern has $p = q$, the most common example being the decussate pattern, when $p = q = 2$, $J=2$ and $N = 4$.

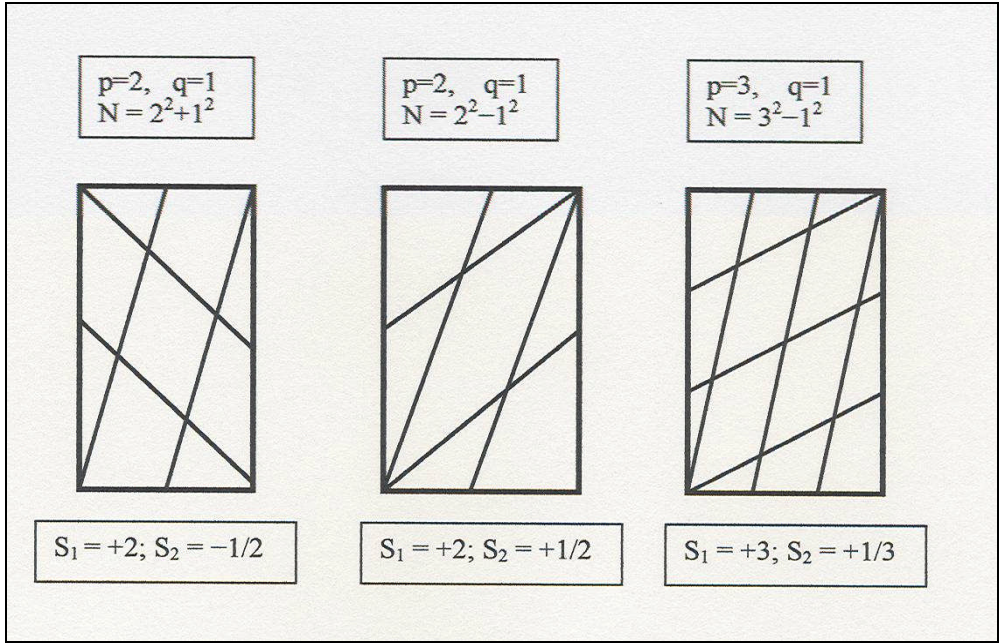

**Figure 5**

**Figure 5:** **Plant Patterns on a Stem**
The present model gives the great preponderance of known plant patterns. Figure 5 shows construction of the Fibonacci patterns for $N = 3$, 5 and 8. A new pair of integers p, q are seen to underlie the more familiar pair 'm, n'. When $p > q > 0$ and p and q are relatively primed, the Fibonacci series emerges. The 'S' represents the slopes of lines where two 'Asps' are equal.

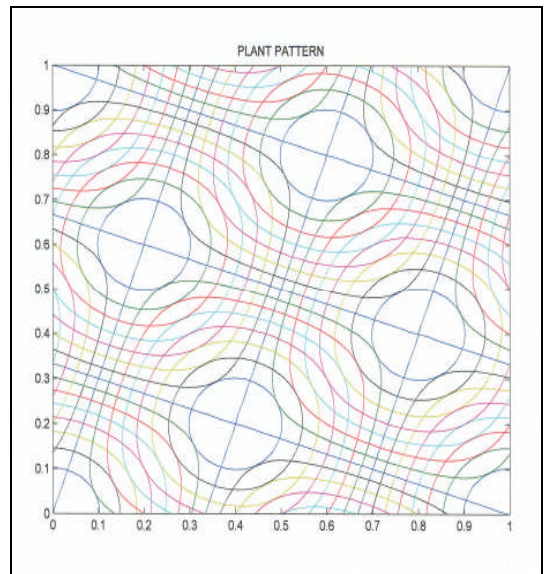

**Figure 6**

**Figure 6**: A contour plot of the Asps of Appendix C shown for the case of $p = 2$, $q = 1$, and $N = 5$, a Fibonacci pattern. The circular regions are the origin of proximo-distal outgrowths. A "master regulatory" gene analogous to Dll of animals is predicted to be found at these centers. The centers

are the intersections of lines where two Asps are equal. Such intersections in animals can possibly precisely position eyes, antennae, limbs, wings, gills, etc., and in plants they position stem outgrowth from the stalk.

Figure 5 shows construction of three patterns, for the case of N = 3, N = 5 and N = 8. Shown below each pattern are the corresponding slopes of the lines. In each case, the two slopes $S_{1,2}$ are (p/q) and (±q/p). The 'leafs' occur at the intersection of the straight lines, and at the origin. These are the intersections of two Margin cell regions as before. Figure 6 shows a contour plot from the equations of Appendix C for the case of p = 2, q = 1, with $N = 5 = 2^2+1^2$.

In the case of the spiral with one leaf per level, growth considered as transition from one Fibonacci pattern to the next occurs simply by adding one leaf to each existing row. The plus/minus sign then alternates in N of eq. (4.3), and p or q is alternately increased to the next value, as shown in Table 1. There are $R_+ = p^2 - (p-q)^2$ pattern rows in the case of the plus sign, and $R_- = q^2 + (p-q)^2$ rows in the case of the minus sign in the expression for N in eq.(5.3). The m's correspond to the usual designation of parastichies, where N = m + n. Figure 5 illustrates three Fibonacci pattern spirals, $N = 3 = 2^2 - 1^2 = 2 + 1$, and $N = 5 = 2^2 + 1^2 = 3 + 2$, and $N = 8 = 3^2 - 1^2 = 5 + 3$.

Whorl patterns have p = q, so that N = 2p, and J = p. Superposed whorl patterns have p > q = 0, when one set of lines has slope zero, and the other set has infinite slope. The common superposed whorl with p = 2, q = 0, and N = p, has two leafs on the same level (J = p = 2) displaced by 180°, with a superposed pair directly above. Compound leaves usually display such a pattern.

There is a simple analytical transformation taking the unit square into an annulus, while preserving the form of the model equations. The coordinate 'y' is mapped into the polar coordinate 'r' while 'x' maps into the polar angle 'θ'. The lines y = 0 and y = $y_o$ are then mapped into two concentric circles, while the lines x = 0 and x = $x_o$ are mapped into the straight lines representing the angles 0 and 2π in the plane. In the case that x maps into an angle less than 2π, the square maps into a conical figure rather than an annulus. The two sets of intersecting straight lines of eqs. (4.1) and (4.2) are in either case mapped into two sets of intersecting logarithmic spirals.

What is clear from eq. (4.3) is that the number '**four**' is not included among spiral patterns, those with a single leaf on a level, according to the model. Such is also very rare in nature. No adaptive or selective reason for this is forthcoming. Rather, it is to be though of as a result of the particular pattern formation algorithm of the present paper.

The Table 1 shows increasing Fibonacci pattern numbers determined as an alternating increase in the two basic determining parameters 'p' and 'q'. The more usual parastichy numbers 'n' and 'm' are shown on the right of the Table.

| N  | $p^2 \pm q^2$ | m+n   |
|----|---------------|-------|
| 2  | $1^2+1^2$     | 1+1   |
| 3  | $2^2-1^2$     | 2+1   |
| 5  | $2^2+1^2$     | 3+2   |
| 8  | $3^2-1^2$     | 5+3   |
| 13 | $3^2+2^2$     | 8+5   |
| 21 | $5^2-2^2$     | 13+8  |
| 34 | $5^2+3^2$     | 21+13 |
| 55 | $8^2-3^2$     | 34+21 |
| 89 | $8^2+5^2$     | 55+34 |
| •  |               |       |
| •  |               |       |
| •  |               |       |

Table 1

**Table 1:**          **The Fibonacci Pattern**

The table shows that as growth occurs, progression from one Fibonacci pattern to the next comes about by addition of one 'leaf' to each row, as space allows. There are $R_- = q^2+(p-q)^2$ rows in the case of the negative sign in $N = p^2 \pm q^2$, when the two sets of straight lines have the same slopes, and $R_+ = p^2 - (p-q)^2$ rows in the case that the two sets have opposite slopes, and illustrated in Figure 4. The relatively primed integers p and q increase alternately with increasing area. The pair (p, q) underlie the more usual 'parastichy' pattern designations (m, n) shown in the right-most column.

### 5. Summary

The present argues for pattern formation agents different from the usual. The concept of 'morphogen' in the usual meaning requires a long-range diffusing substance. The term '**Asp**' is used here to refer to the density of a particular activated signalling pathway, and thus to the density of its associated factor that activates its specific transcription factor. The term 'Asp' denotes a spatial region in which specific transcription factors have been activated in the nucleus, and where particular selector genes are activated (Gerhart and Kirschner, 1997). As such, no discussion of whether ligand diffuses around cells or through them is required, nor is there discussion of the method of achieving long range diffusion.

Section 2 shows how the interaction of two signaling pathways, acting as modulators or patterning agents of developmental genes, can interact with morphogenetic movement. Specific genes will give rise to specific cell shapes. Such epithelial sheet movements will necessarily act back to affect gene activation in turn, via the Lagrangian. The two signaling pathways will rise spontaneously from zero amplitude when available area and geometry permits, followed by decay when the amplitude reaches a specified level. Then a subsequent pair, consisting in general of two new 'Asps', arises when the geometry is appropriate, providing a binary tree of unique Asp gene activation.

The origin of multicellulars is proposed in Section 2 to arise at the Pre-Cambrian as the simultaneous interaction of cadherins (Abedin and King, 2008) and the Wnt pathway, along with a second pathway, (e.g., especially Hedgehog (Hh)) (Nusse, 2003). Cadherins are important players in cell-cell adhesion and cell structure. The adaptive process of gastrulation then occurs as a simple and straightforward interaction of the patterning of the two interacting pathways with the closed epithelial sheet of polarized cells. Two points are emphasized as being crucial to a pattern model. The first is that no pattern may arise (from zero amplitude, as at present) until a sufficient total area (or radius) of the blastula is achieved, as is observed. Secondly, it is crucial that any model of pattern formation specify stem cell 'niche(s)', since stem cells are not distributed in space at random. The present model does not specify such patterning.

Growth is assumed to occur from Margin cell regions. The intersection of 'linear' Margin cell regions is posited as giving precise locations for the outgrowth for such appendages as legs, gills and wings, as well as specifying positions for eyes. The surrounding activated signaling pathway ('Asp') pair in any growth cycle regulates the Margin cells. It is emphasized that pattern models must attempt to provide rather precise information regarding leg, gill and wing outgrowth placement, and also placement of eyes and antennae. In the present model, plants and animals display analogous behavior in the placement of 'leaf outgrowths' and animal appendages.

A model of segment formation is discussed, with emphasis on the positional placement of legs, gills and wings at Margin cell intersections. Growth occurs from the posterior of animals, since it is known that the order of genes in each Hox cluster along a chromosome (from one cluster in primitive animals to four in vertebrates) correspond to the body region in which they are expressed, i.e., anterior to posterior along the body axis.

Also based on the model, plant patterning of 'leafs' on a stem provides for a simple algorithm to give all of the main plant patterns. The number 'four' is conspicuously absent in spiral plant patterns, patterns where there is only one leaf per level, a result not explainable on the basis of natural selection. Plant patterning here assumes that plants have the same basic patterning mechanism as animals, although with very different biochemistry. A prediction is that a "master regulatory" gene will be found at each point of 'leaf' outgrowth in all (or most) plants, in analogy to the occurrence of such 'master genes' as Distall-less (Dll) in outgrowths from animals.

The important question of the patterning of stem cells is not discussed. This area of stem cells is too fraught with mystery in general at present, although the suspicion is that it is somehow related to the Margin region. But how? Could the combination of two transcription factors conspire to specify a stem cell?

## Appendix A

The model consists of two interacting signaling pathways. There are two simple elements of the model. Any model containing these two simple elements will produce interesting patterning, so that the addition of further 'bells and whistles' will not alter the basic concept. The purpose of this appendix is to outline the mathematical model.

Attention is focused on a small cluster of cells, ~ five-ten, when use of such terms as 'ligand density' and 'receptor density' has meaning. The cells are to be thought of as comprising a closed epithelial surface, and the densities of the model have dimensions of 'number/area'. Variation of the Asps (the R's or L's) along the apical-basal cell direction is not considered, or rather thought of as being an averaged value in this dimension.

First of all, each such 'cell', or rather cell cluster (~5 or 10 cells), produces ligand proportional to the level of receptor activation of like kind in a given small region. Asp $R_1$, an activated receptor, stimulates production of $L_1$, otherwise the process would be limited to a purely local process in the absence of "like-ligand" production, with the particular cell in question then acting as a ligand 'sink'.

The second key element in the model is that activation of a pathway acts to inactivate the other; as $R_1$ increases, the level of ligand production $L_2$ is decreased, and similarly for $R_2$. This may occur in a number of ways; the simplest to imagine is that one emitted ligand may block the receptor of the second type.

The equations representing such a process are then able to be written at once, and are

$$\frac{\partial}{\partial t} L_1 = D_1 \nabla^2 L_1 + \alpha R_1 - \beta R_2 + NL., \qquad (A.1)$$

$$\frac{\partial}{\partial t} L_2 = D_2 \nabla^2 L_2 + \beta R_2 - \alpha R_1 + NL., \qquad (A.2)$$

$$\frac{\partial}{\partial t} R_1 = C_1 \overline{R}_1 L_1 - \mu R_1, \qquad (A.3)$$

$$\frac{\partial}{\partial t} R_2 = C_2 \overline{R}_2 L_2 - \nu R_2. \qquad (A.4)$$

The first two terms in eqns. (A.1), (A.2) represent in the usual way random diffusion of the ligands in the extracellular space. All parameters in the model (e.g., α, β, $D_1$, $C_1$, μ, ν) have positive values, as do also, of course, the densities $L_1$, $L_2$, $R_1$ and $R_2$. The terms $\alpha R_1$ in eqn.(A.1) and $\beta R_2$ in eqn.(A.2) represent the production of 'like' ligand by the corresponding activated receptor. These same terms are used to represent the fact that activation of receptors of density $R_2$ deactivate or turn off production of free ligands of density $L_1$, and vice versa. A region of high activation of one Asp then implies low activation of the second. The term NL on the r/h/s of eqns. (A.1) and (A.2) indicate that there are expected to be nonlinearities; saturation effects set in for large enough values of either active receptor density.

In the numerical work, (e.g., Figure 1) the non-linearity is introduced by letting the difference of the two Asps go to an expression that decreases as the Asp level increases, that is,

$$\alpha R_1 - \beta R_2 \rightarrow (\alpha R_1 - \beta R_2)/(1 + ((\alpha R_1 - \beta R_2)/c)^2), \qquad (c \sim 1).$$

The transmembrane receptors, which reside in the lateral cell plasma membrane, are relatively immobile. The respective activated densities decay at rates μ and ν, and this 'decay' returns the receptors to their inactive state. Two first terms on the right side of eqns.(A.3), (A.4) say that there is a positive rate of change of $R_1$ or $R_2$ proportional to both the density of empty receptor sites ($\overline{R}_1, \overline{R}_2$) and also to the density of free ligands at the particular local cell site. The density of empty sites may be obtained from the expression

$$R_1 + \bar{R}_1 = R_o + \eta R_1, \qquad (R_0 = const.),$$

where the last term on the r/h/s expresses the possibility that the total number of receptors of each type (e.g., '1') increases with activation of that same type receptor, and new (empty) receptors are thus added. Then the empty receptor site density may be written

$$\bar{R}_1 = R_o(1 - \varepsilon_1 R_1), \qquad (0 < \varepsilon_1 \leq 1, \varepsilon_1 \equiv (1-\eta)/R_o), \qquad (A.5)$$

and similarly for type '2'. The values $\varepsilon = 1$ (and $\eta = 0$), implies that there is no receptor augmentation $\sim R_1$, while $\varepsilon \sim 0$ implies either that there is a new empty receptor created for (almost) every one occupied, or that there are very many more empty sites than occupied ones. When eqn. (A.5) and the analogous equation for type '2' is used in eqns. (A.3) and (A.4) to eliminate the unoccupied site densities, the model then comprises four coupled equations for four unknowns. The coupling from epithelial shape to Asp, and back is discussed elsewhere (Cummings, 2005, 2006).

The small amplitude, time independent ($\partial/\partial t = 0$) version of eqns. (A1)-(A5) are simply the Helmholtz and Laplace equations

$$\nabla^2(R_1 - fR_2) + k^2(R_1 - fR_2) = 0, \qquad (A.6)$$

and

$$\nabla^2(R_1/k_1^2 + fR_2/k_2^2) = 0. \qquad (A.7)$$

The definitions $k^2 = k_1^2 + k_2^2$, $f = \beta/\alpha$, $k_1^2 = \alpha C_1 R_0/(D_1\mu)$, and $k_2^2 = \beta C_2 R_0/(D_2\nu)$ have been used. The k's are the two inverse lengths of the model.

It is easily shown that no solution to eq. (A.6) exists until a sphere (e.g.) reaches a particular radius, say $R_0$. The condition that a solutions for $R_1$, $R_2$ can just begin to emerge on the sphere is that

$$(k_1^2 + k_2^2)/2 = 1/R_0^{-2}.$$

Below this sphere (blastula) radius, the **Asps ($R_1$ and $R_2$) remain zero.**

Several forms may serve to model the 'N.L' terms on the r/h/s of eqns. (A.1) and (A.2). The model does not appear to be very sensitive to this choice, or even if they are zero. The simplest, and the one used in present simulations is to let $R_1 - (\beta/\alpha)R_2 \to (R_1 - (\beta/\alpha)R_2)/(1 + ((R_1 - (\beta/\alpha)R_2)/c)^2)$.

The constant 'c' is ~1. Others forms will no doubt lead to other surface shapes for the invagination. The 'Asps' of the present work are taken as

$$\Phi_1 = R_1/R_o, \qquad \Phi_2 = (\beta/\alpha)R_2/R_o. \qquad (A.8)$$

Thus the 'Asps' denote the density of activated signalling pathway. One possible process of producing ligand upon activation of the cell surface receptor could involves numerous steps, involving (e.g.) gene transcription, the endoplasmic reticulum (ER), the Golgi complex, and finally perhaps secretion from the cell. This time is expected to be considerable compared to the time for a free ligand in a given spatial region to become attached to its receptor and to activate the pathway. However, it is supposed here instead that $R_{1,2}$ acts downstream to release already stored ligand, by a route that bypasses the nucleus. Such ligands are supposed stored at, e.g., a constant rate by an unspecified cellular mechanism. The cell maintains a relatively constant store of ligand awaiting a release signal ~ R (analogous (in this respect only!) to the situation of neurotransmitters in neurons). The two times **a**: emission time interval between receptor activation and like ligand emission, and **b**: empty receptor uptake of ligand L) can thus be comparable. This is the situation envisioned here, and will have to serve as a prediction of the model at this point: the activated receptor $R_{1,2}$ releases ligand already stored in vesicles, so that this time is appreciably shorter than ligand production and storage via gene activation, ER and Golgi. This provides for a 'pre-pattern'. Importantly for the model, Wnt has two known modes of action, one that bypasses the nucleus, and a second (and most discussed) 'canonical' pathway leading to gene activation via stabilization of nuclear β-catenin. The former 'non-canonical' path bypassing the nucleus acts possibly (at least in part) to release stored Wnt ligand relatively rapidly through exocytosis into the extracellular space.

The coupling of the Asps to patterning (Cummings, 2005, 2006) is through the (exact) Gauss (~1800) equation, namely

$$\nabla^2 (\ln_e(g(u,v)) = -2K(R_1, R_2) \quad . \tag{A.8}$$

The operator $\nabla^2$ in eq. (A.8) contains the geometry of the surface, as also in the same operators in the equations (A.1) to (A.7) above. In terms of the most convenient coordinates as used in the numerical work here, namely the conformal coordinates 'u, v' (Cummings, 2005, 2007) when there is only one metric component g(u,v), the 'Laplace-Beltrami' operator of eq. (A.8) takes the form

$$\nabla^2 = \frac{\partial^2/\partial u^2 + \partial^2/\partial v^2}{g(u,v)}. \tag{A.9}$$

Thus there are three coupled second order partial differential equations to be numerically solved in general, in steady state. The case of axial symmetry as carried out here reduces the numerical difficulty immensely.

The dependence of the Gauss curvature K (on the right hand side of eq.(A.8)) on the two 'Asps' $R_1$ and $R_2$ turns out to be surprisingly straightforward (Cummings, 2005, 2006) due to the constraints and invariance which are imposed on it. The mean curvature H is taken, as a model, to be of one sign (plus), when apical cell area is larger than basal at a locus on the middle surface of the epithelial sheet, and $R_1$ large, and secondly, as the opposite sign (negative) when local basal cell area is larger than apical and $R_2$ is large, in the same small region. By definition, $K \leq H^2$, and the two are equal only when they are both locally describe spherical surfaces. The invariance of K means that it is independent of coordinates, and then requires that the negative part of K ($K \equiv H^2 - D^2$) depends on the gradient squared of the difference of Asp densities (Cummings, 2005). Then Gauss curvature K has been taken to be of the two parameter form

$$K = (4\pi/A)(1+\lambda_1(R_1-R_2))^2 - (\lambda_2 \text{grad}(R_1-R_2))^2. \tag{A.10}$$

The invariant expression "(grad(y))²" is, in conformal coordinates,

$$(\text{grad } (y))^2 = [(\partial y/\partial u)^2 + (\partial y/\partial v)^2]/g(u,v).$$

The constants $\lambda_1$ and $\lambda_2$, pure numbers, have values in the numerical work of Figure 1 of '7' and '10' respectively. These are not very specific; the model is "robust". A gradient is required in (A.10) because the "twist" property of the cell region requires the specification of a direction (Cummings, 2005). It is to be emphasized that a negative region of K must exist in order for gastrulation to occur, and that the Gauss-Bonnet integral over the endoderm be zero.

## Appendix B

The linear equations are (from eqs. (A.6), (A.7) and (A.8) ) are

$$\nabla^2 (\Phi_1 - \Phi_2) + k^2 (\Phi_1 - \Phi_2) = 0 \qquad (B.1)$$

$$\nabla^2 \left( \frac{\Phi_1}{k_1^2} + \frac{\Phi_2}{k_2^2} \right) = 0. \qquad (B.2)$$

The constant parameters k, $k_1$, $k_2$ and $f \equiv \beta/\alpha \sim 1$ have been defined in Appendix A, and $k^2 = k_1^2 + k_2^2$. The solution to the two equations (B.1) and (B.2) are given in the case of cylindrical symmetry, but now when the topology is that of a torus, or donut. All solutions must be periodic in both coordinates, z and φ, and this requires that

$$\Phi_1 = C + Dk_1^2 Cos(\pi mz/L) \cdot Cos(n\phi), \qquad (B.3)$$

and

$$\Phi_2 = C - Dk_2^2 Cos(\pi mz/L) \cdot Cos(n\phi). \qquad (B.4)$$

The constant $Dk_2^2 < C$. Here two integers 'm' and 'n' are defined, and the solution is periodic for $z \to z + 2L$ as well as $\varphi \to \varphi + 2\pi$ (see Figure 2). The condition from eq. (B.1) relating R, L, m and n for the small amplitude solution of eqns. (B.3) and (B.4) is

$$1 = (\pi m/kL)^2 + (n/kR)^2. \qquad (B.5)$$

Equation (B.5) is eq. (3.1) of the text. What is clear from eq. (B.5) is that while kR remains below unity, n must remain zero. So it is supposed that $kL = m\pi$ for some m while $kR < 1$. As kR passes unity and becomes $\approx 2^{1/2}$, n now becomes unity, and kL also grows to become $kL \approx 2^{1/2} m\pi$. At this point, a through-gut, bilateral animal has emerged, one with m 'segments'. Length L may continue to increase so that $kL \approx 2^{1/2} \pi \cdot m$, where $m \gg 1$, while keeping $kR \approx 2^{1/2}$. This is shown in Figure 4a.

The next growth cycle may occur after a further increase in radius R, until $kR \approx 2^{3/2}$, when now n = 2. Length L may continue to increase so that $kL \approx 2^{1/2} \pi \cdot m$, where $m \gg 1$. A new pattern of small amplitude begins to emerge at each growth cycle, at each area sufficient so that eq. (B.3) can be satisfied. There are now four small linear regions running down the axis where $Cos(n\varphi) \approx 0$, n = 2, where both 'Asps' are nearly equal to the same constant C in eqs. (B.3) and (B.4). This is shown in Figure 4b by the two horizontal lines down the axis. Along with these linear regions, there are also a series of (small, vertical) circular regions, three shown in the Figure 4, representing regions where the two Asps are small and cannot effect cell determination, and denoted as Margin cell regions. The intersections of these two sets of Margin cell regions, vertical and horizontal in Figure 4b provide positional specification for (ventral) limb or dorsal appendage outgrowth. Remembering that each cycle of pattern formation overlays the previous, the appendage position is designated, along with information distinguishing front from back, inside from outside of (e.g.) a leg.

## Appendix C

The patterns that appear on plant stems are given in light of the model. Clearly, what is being proposed is the possibility that, in the broadest terms, the patterning mechanisms of plants is similar to that of animals. Two Asps, presumably one of them being auxin and its receptor, follow the same two conditions as in animals. An intriguing connection has been found between plant gibberellin signaling and Drosophila armadillo (Amador et al., 2001). Again the basic result is a model with two 'Asps' in which the two Asp densities avoid each other (Cummings and Strickland, 1998).

The solutions to eqs. (B.1) and (B.2) are,

$$\Phi_1 = C + Dk_1^2 [Cos(2\pi px/x_o + 2\pi qy/y_o) + Cos(2\pi qx/x_o \pm 2\pi py/y_o)], \quad (C.1)$$

$$\Phi_2 = C - Dk_2^2 [Cos(2\pi px/x_o + 2\pi qy/y_o) + Cos(2\pi qx/x_o \pm 2\pi py/y_o)]. \quad (C.2)$$

Here $Dk_2^2 < C$. The boundary conditions are that the solutions be doubly periodic, periodic in the angular variable '$x/x_0$' around the stem axis, and also in the "repeating" variable along the stem '$y/y_0$'. Here k is an (inverse) length, and satisfies the equation

$$\left(\frac{k}{2\pi}\right)^2 = (p^2 + q^2) \cdot \left(\frac{1}{x_o^2} + \frac{1}{y_o^2}\right). \quad (C.3)$$

Positions of the 'florets' or 'leafs' are presumed to originate from the same source as in the outcroppings (e.g., legs) of the animal case. Again the 'Margin cells' are in regions between the two larger determined regions. The intersection of two Margin cell regions (or lines) is given by the case where $\Phi_1$ of eq. (C.1) is equal to $\Phi_2$ of eq. (C.2). This requires that

$$px/x_0 + qy/y_0 = i/2, \quad (C.4)$$

$$py/y_0 \pm qx/x_0 = j/2. \quad (C.5)$$